# Energy calculations of charged point defects on surfaces

Feifei Li[*], Zhenpeng Hu, Ziping Niu and Lixin Zhang[†]

*School of Physics, Nankai University, Tianjin 300071, People's Republic of China*

## Abstract

We present a virtual ionic crystal (VIC) method to calculate energies of charged point defects on surfaces. No artificial charge but an actual zero-dimensional (0D) species is introduced to charge a defect. Effect of dielectric substrate on lattice energy is depressed through suitable configuration of the unit cell. The lattice energy approximates to Madelung energy with defect and 0D species considered as point charges in vacuum. Energy required to charge the defect is derived from charge quantity on the defect in VIC, energy of unit cell, and energy required to charge the 0D species.

PACS numbers: 73.20.Hb, 71.15.Nc, 32.10.Hq, 71.15.-m

[*]lff@nankai.edu.cn
[†]lxzhang@nankai.edu.cn

Density functional theory (DFT) with periodic boundary conditions (PBC) has been widely used in physics, chemistry and materials science. It is ideal for obtaining properties of neutral systems. When applying it to charged systems, charge neutrality is essential to avoid electrostatic energy divergence. The routine scheme is introducing jellium background as counter-charge, and posteriorly eliminating the excess Coulomb interaction between system charge, counter-charge and their images. For charged homogeneous systems, the posterior corrections are relatively easy and provide quit well results [1, 2]. For charged heterogeneous systems, different dielectrics affect the Coulomb energy according to their dielectric constant, position, size and shape; the posterior corrections are much more complicated and hindered the application of DFT calculation with PBC until recent years [3-8]. A simple and interesting heterogeneous system is ionic doping or adsorption on material's surface. Material substrate and vacuum coexist in a unit cell. Some methods based on jellium counter-charge have been proposed to calculate energies of charged defects on material's surfaces [4-6]. Because jellium fills the unit cell evenly, electric displacement field spreads equally in the vacuum and substrate around the defect. The substrate's effect on Coulomb energy is commonly not negligible, and the corresponding corrections are sophisticated and specific for given substrate, let alone determination of the dielectric constant in advance.

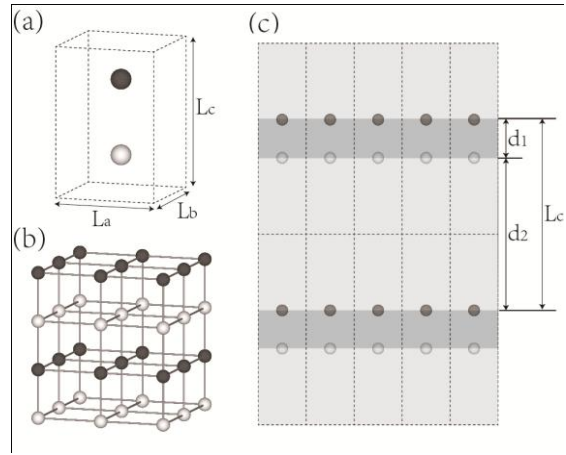

FIG.1 (a) A positive and a negative point charge in a unit cell. (b) The charges on $3 \times 3 \times 2$ lattice points of VIC. (c) (100) plane of VIC. Space of VIC is divided into intra double layer (dark grey) and inter double layer (light grey) spaces.

Using point counter-charge is an alternative solution for charged surface defect, which can avoid consideration of substrate's effect on the Coulomb energy. In principle, at a long-distance limit, a charged defect acts as a point charge. Using three-dimensional PBC, putting a point charge and its point counter-charge in a unit cell, there is a VIC with inverse charges arranged in alternate layers, as sketched in Fig. 1 (a), (b). Adjusting the configuration of the unit cell as Fig. 1 (c), the VIC takes the form of a series of double layers. When thickness of the interspace $d_2$ is much larger than that of the double layer $d_1$, the Coulomb energy is mainly due to electric field intra the double layers. Existence of dielectric in the interspace affects the energy weakly. Then, lattice potential energy of a point charge with substrate approximates to its Madelung energy in free space, and is easily managed. Inferior to jellium, point counter-charge may share electrons with defect during self-consistent DFT calculation and make fractional occupation on defect deviated from reality. Once the relationship between energy and electron occupation of a defect is obtained, energy of the defect with integer multiple of unit charge can be derived.

In this manuscript, we introduce an actual 0D species to compensate a charged surface defect in a unit cell, which keeps the unit cell neutral. Energy of the unit cell and charge quantity transfered between the defect and the 0D species are derived from self-consistent calculation on VIC. From the energy of unit cell and transfered charge quantity in VIC, energy required to charge a defect is obtained after Madelung energy correction. Carbon substituting nitrogen and boron atoms ($C_N$ and $C_B$) on single layer h-BN sheet, and chlorine vacancies ($V_{Cl}$) on NaCl bilayer sheet are taken as examples.

As usual, a surface defect is modeled by a defect on portion of a substrate in a unit cell. There are vacuum intervals between the periodically repeated substrates perpendicular to extending directions. Putting a 0D species into vacuum interval of a unit cell and keeping it well separated with substrate, the 0D species and substrate own distinct electron density. In light of variational principle for DFT, there is electron density redistribution to lower the total energy. Change of electron density on the substrate is locally around defect. One of the 0D species or defect acts as a donor, another one acts as an acceptor. Their electron density changes relative to respective neutral state are denoted $\Delta n_d$ and $\Delta n_a$. Integrations of $\Delta n_d$ and $\Delta n_a$ are donor and acceptor's charge quantities $\pm q$. A VIC forms with ions possessing charge $\pm q$ on the lattice points.

Electron energy of an ion in the VIC is $T + V^{nucl} + U + V^{latt}$, where $T$ is kinetic energy, $V^{nucl}$ is potential energy from its nuclei potential, $U$ is energy of electron-electron interaction inner the ion, $V^{latt}$ is lattice energy from lattice potential induced by other ions in VIC. The VIC's energy per cell is energy sum of donor and acceptor:

$$E_{cell} = \{T_d + V_d^{nucl} + U_d\} + \{T_a + V_a^{nucl} + U_a\} + \frac{1}{2}\{V_d^{latt} + V_a^{latt}\} \quad (1)$$

In the long-distance approximation, lattice potential through each ion is approximately constant and proportional to $q$. Then lattice energies in last brace of Eq. (1) are related to integrations of $\Delta n_d$ and $\Delta n_a$, but free from their spatial distributions. Self-energies in the first two braces are functional of $\Delta n_d$ or $\Delta n_a$ respectively. Electron density integration of an ion is concerned with unit cell's configurations, and its electron distribution is determined by its nuclei potential. For given donor and acceptor, $\Delta n_d$ and $\Delta n_a$ are unique to $q$, and each term in Eq. (1) is function of $q$.

We denote $\Delta E_d(q)$ and $\Delta E_a(-q)$ self-energy changes of donor and acceptor from neutral state to $\pm q$ charged state. The values $\Delta E_d(+e)$ and $\Delta E_a(-e)$ are ionization energy (IE) of donor and additive inverse of electron affinity (EA) of acceptor relative to infinity with zero potential. Subtracting energy sum of neutral donor and acceptor from $E_{cell}$, Eq. (1) is rewritten as

$$\Delta E_{cell} = \Delta E_d(q) + \Delta E_a(-q) + Pq^2 \quad (2)$$

where $P$ is parameter decided by configurations of the unit cell. Configuring the VIC as Fig.1 (c) with substrate in the interspace, there is $P \approx E_M/Q^2$, where $E_M$ is the Madelung energy per cell with point charges $Q$ and $-Q$ at donor and acceptor's centers. Unlike actual crystal, the VIC's $E_M$ may be positive.

Ideally, when shared electrons are limited to each highest energy level of donor and acceptor, self-energy of the donor or acceptor linearly changes with its electron occupation. However, the relation between self-energy and electron occupation significantly deviate from linearity, due to electron-electron interaction, and the unphysical self-interaction of electrons. We limit the shared electrons to highest energy level of donor and acceptor, and expand $\Delta E_d(q)$ and $\Delta E_a(-q)$ to power series of $q$

$$\left.\begin{array}{l}\Delta E_d(q) = \alpha_d + \beta_d q + \gamma_d q^2 + \sigma_d \\ \Delta E_a(-q) = \alpha_a - \beta_a q + \gamma_a q^2 + \sigma_a\end{array}\right\} \quad (3)$$

Theoretically, the coefficients $\alpha$ should be zero and discarded. Besides approximation in calculation,

their appearances are because of two reasons. First, charged donor and acceptor frequently have different geometries from that in neutral state. In VIC calculations, their geometries should keep in aimed ionic state. Second, calculation configurations of VIC may be different from that of neutral donor and acceptor, for instance, spin polarization considered or not. When the two differences exist, $\beta$ and $\gamma$ are also impacted, and the coefficients are valid for selected ionic state only. The terms with higher power are lumped into $\sigma$. We expect that the components $\sigma$ could be ignored, and test it later.

In ground state, $\Delta E_{cell}$ has minimum relative to variations of electron density or charge in a specific volume. Taking donor as the specific volume, there is

$$\frac{\partial}{\partial q}\Delta E_{cell} = \frac{\partial}{\partial q}[\Delta E_d(q) + \Delta E_a(-q) + Pq^2] = 0 \qquad (4)$$

Expanding $\Delta E_d(q)$ and $\Delta E_a(-q)$ as Eq. (3):

$$(\beta_d - \beta_a) + 2(\gamma_d + \gamma_a)q + \frac{\partial}{\partial q}(\sigma_d + \sigma_a) + 2Pq = 0 \qquad (5)$$

Dropping all $\sigma$, then

$$\gamma_d + \gamma_a + P = -\frac{(\beta_d - \beta_a)}{2q} \qquad (6)$$

And

$$\Delta E_{cell} = (\alpha_d + \alpha_a) + \frac{(\beta_d - \beta_a)}{2}q \qquad (7)$$

The $\Delta E_{cell}$ changes linearly with $q$. Magnitude of $\sigma$ is estimated by degree of the linearity. The $\alpha_d + \alpha_a$ and $\beta_d - \beta_a$ are obtained from the intercept and slope. Then $\gamma_d + \gamma_a + P$ can be deduced from Eq. (6), $\gamma_d + \gamma_a$ is achieved by eliminating $P$ with Madelung energy. During the VIC calculations, long distances between the ions have to be maintained to make the charged places interacting as point charges. In addition, thickness of inter double layer spaces $d_2$ should be much larger than intra double layer $d_1$ and substrate's thickness to depress the influence of dielectric substrate on lattice energy.

After obtainment of the coefficients, energy required to transfer $q$ between isolated donor and acceptor is

$$E_{trs}(q) = \Delta E_d(q) + \Delta E_a(-q) = (\alpha_d + \alpha_a) + (\beta_d - \beta_a)q + (\gamma_d + \gamma_a)q^2 \qquad (8)$$

Substituting $q$ with unit charge, $E_{trs}(e)$ is difference between donor's IE and acceptor's EA. Since IE or EA of 0D species are easily experimentally determined or calculated in jellium scheme, EA or IE of a surface defect is achieved after $E_{trs}(e)$.

The DFT calculations were carried out on plane wave basis set. Exchange and correlations were treated with the Perdew-Burke-Ernzerhof functional [9] implemented in the VASP code [10]. The plane-wave cutoff was set to 400 eV. Carbon substituting nitrogen and boron atoms on single layer h-BN sheet acted as acceptor and donor respectively. Chlorine vacancy on bilayer NaCl sheet acted as donor. Chlorine and potassium atoms were taken as 0D donor and acceptor. Lateral ranges of h-BN were from 4×4 to 9×9 primitive cells, that of NaCl were from 2×2 to 5×5 primitive cells. The geometries of +e or −e charged defects were obtained by relaxation in jellium background until all residual forces smaller than 0.02 eV/Å. Calculations on VIC were performed without spin polarization in consideration of only paired electrons exiting in the ±e charged state. Spin polarized calculations were taken on the energies of neutral donor and acceptor. The vertical size $L_c$ was 75 Å. Distance between donor and acceptor varied from 11 to 20 Å. K-point mesh for cells containing only atoms was gamma only, for cells containing h-BN and NaCl are 8×8×1 and 5×5×1 or higher corresponding to their primitive cell respectively. Charge quantities were estimated within a Bader method[11]. The Madelung energy was obtained from VESTA [12].

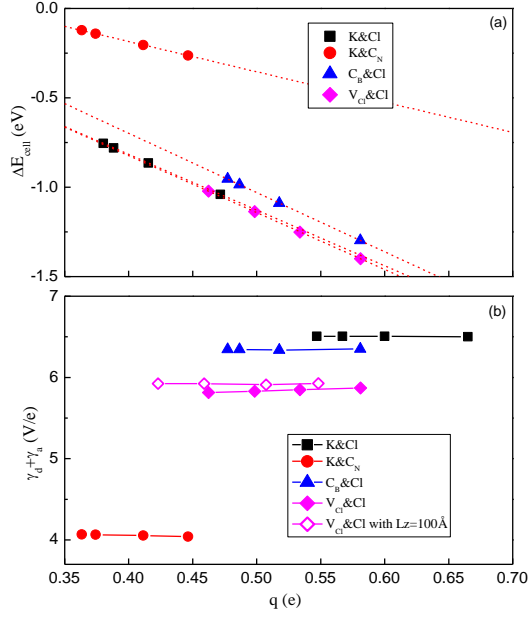

FIG.2 (Color online). Dependences of $\Delta E_{cell}$ and $\gamma_d + \gamma_a$ on $q$ for some donor & acceptor couples. The $C_N$ and $C_B$ are on 9×9 h-BN sheet, $V_{Cl}$ is on 3×3 bilayer NaCl sheet. $\Delta E_{cell}$ and $q$ are lineally fitted with short dash lines in (a). The $\gamma_d + \gamma_a$ in (b) is Madelung energy corrected value from $\gamma_d + \gamma_a + P$. The extra data denoted with open diamond in (b) are for $V_{Cl}$ & Cl with $L_c$=100 Å.

In this work, dependence of $\Delta E_{cell}$ on $q$ is derived from altering distance between donor and acceptor. Through our calculations, dependences of $\Delta E_{cell}$ on $q$ for all donor & acceptor couples are well linearly fitted with standard errors of intercept and slope less than 0.05 eV and 0.05 V, which demonstrate that ignoring σ is acceptable with general precision. Some of the fitting curves are shown in Fig.2 (a). The intercept and slope values are concerned with $\alpha_d + \alpha_a$ and $\beta_d - \beta_a$ as Eq. (7). Then $\gamma_d + \gamma_a + P$ are calculated with Eq. (6), and $\gamma_d + \gamma_a$ are achieved by elimination of parameters $P$. The data presented in Fig.2 (b) show that the $\gamma_d + \gamma_a$ are nearly irrespective of $q$, in other word, nearly irrespective of the distance between donor and acceptor. It indicates that the lattice energy approximates well to Madelung energy of point charge, and setting $L_c$ to 75 Å is tolerable for the present cases. Increasing $L_c$ to 100 Å improves the approximation about $V_{Cl}$, but it is not essential. The coefficients and corresponding $E_{trs}(e)$ are presented in Table. I.

TABLE I. Calculated results of donor & acceptor couples in Fig. 2. $L_a$ are lateral dimensions of unit cells. Coefficients $\alpha_d + \alpha_a$ and $\beta_d - \beta_a$ are derived from dependences of $\Delta E_{cell}$ on $q$ in Fig. 2 (a), $\gamma_d + \gamma_a$ are values at respective largest $q$ in Fig. 2 (b). $E_{trs}(e)$ are calculated from the coefficients with Eq. (8).

|  | K&Cl | K&$C_N$ | $C_B$&Cl | $V_{Cl}$&Cl |
|---|---|---|---|---|
| $L_a$ (Å) | 20.00 | 22.62 | 22.62 | 17.05 |
| $\alpha_d + \alpha_a$ (eV) | 0.43 | 0.65 | 0.55 | 0.65 |
| $\beta_d - \beta_a$ (V) | -6.23 | -3.39 | -6.63 | -6.37 |
| $\gamma_d + \gamma_a$ (V/e) | 6.73 | 4.07 | 6.35 | 5.87 |
| $E_{trs}(e)$ (eV) | 0.93 | 1.32 | 0.27 | 0.15 |

Generally, defect in charged state produces much wider impact on the substrate's geometry and electron structure than neutral state. Size of substrate in the unit cell significantly influences calculated properties of the charged defect. We uniformly enlarge lateral size of the substrates, and plot the dependences of $E_{trs}(e)$ on $1/L_a$ in Fig. 3. The charged defects with relaxed geometry have obvious size effect. Fixing the geometry to bulk reduces the size effect of $V_{Cl}$ notably. Since charged defects are commonly sparse on substrates, modeling a charged defect with a small portion of substrate usually results in large deviation from reality. We linearly fit the dependences of $E_{trs}(e)$ on $1/L_a$ and extrapolate $E_{trs}(e)$ to $E_{trs}^{\infty}(e)$ at infinite large $L_a$. The $E_{trs}^{\infty}(e)$ are considered as energies required to transfer an electron between real defects and respective 0D species. Calculated IE of K and EA of Cl by us in jellium scheme are 4.51 and 3.72 eV. The $E_{trs}^{\infty}(e)$ of K & Cl couple is 0.88eV, agrees with the difference between IE of K and EA of Cl. Taking K and Cl as benchmark, from $E_{trs}^{\infty}(e)$ of the donor & acceptor couples, EA of $C_N$ are 3.83 eV, IE of $C_B$ are 3.20 eV. The IE of $V_{Cl}$ is 3.25 eV; IE of $V_{Cl}$ with fixed bulk positions is 4.26 eV.

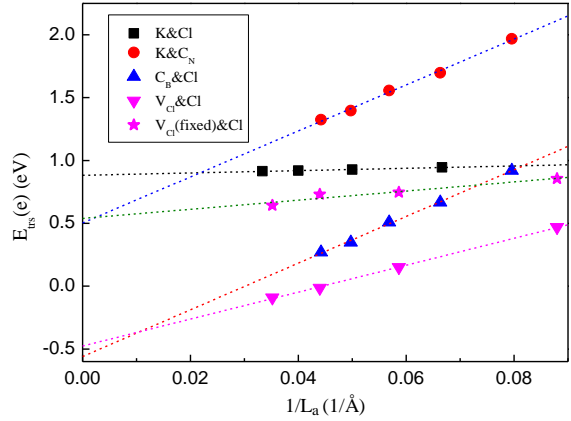

FIG. 3 Dependence of $E_{trs}(e)$ on $1/L_a$ and linear fit with short dash lines. The intercepts are $E_{trs}^{\infty}(e)$. Data indicated with stars are about $V_{Cl}$ on bilayer NaCl sheet with atoms fixed at their bulk positions.

In solid-state physics, substrate's valence band maximum (VBM) commonly acts as electron's reference level. Our calculated VBM relative to vacuum is -5.79 eV for h-BN, -6.54 eV for NaCl bilayer and -6.17 eV for NaCl bilayer with fixed bulk positions. Thermodynamic transition level $\epsilon^{q1/q2}$ is defined as the Fermi-level position where charge states $q1$ and $q2$ have equal energy, as Ref. [13]. The $\epsilon^{+/0}$ of $C_B$ and $\epsilon^{0/-}$ of $C_N$ from VIC method agree well with that from jellium method in literatures, as listed in Table II. The $\epsilon^{+/0}$ of $V_{Cl}$ depends significantly on geometry of the bilayer, it varies from 3.29 to 1.90 eV according to whether the geometry is relaxed or not. That is in good agreement with the experimental result in Ref. [14], where the unoccupied state energies of $V_{Cl}$ at a NaCl bilayer on Cu (311), (100) and (111) support are 2.03, 2.57, and 2.83 eV. Taking chemical potential of Cl from isolated $Cl_2$ molecule and setting Fermi energy at VBM, formation energy of $V_{Cl}^+$ with fixed bulk geometry is 1.50 eV in bulk, and increases to 1.89 eV on bulk's surface [4]. It further increases to 2.46 eV on bilayer sheet from our calculation.

TABLE II. $\epsilon^{0/-}$ and $\epsilon^{+/0}$ of surface defects. Data are calculated from VIC method and cited from literatures.

|  | $\epsilon^{0/-}$ (eV) | $\epsilon^{+/0}$ (eV) | | |
| --- | --- | --- | --- | --- |
|  | $C_N$ | $C_B$ | $V_{Cl}$ | $V_{Cl}$ (fixed) |
| VIC | 1.96 | 2.59 | 3.29 | 1.90 |
| Literatures | 2.03[a], 1.86[b] | 2.43[a], 2.50[b] | 2.83—2.03[c] | |

[a]Ref. [5].
[b]Ref. [6].
[c]Ref. [14].

In summary, we introduced a VIC method to calculate energy of charged point defect on a material's surface. An actual 0D species was employed to charge a defect in a unit cell. Through suitable configuration of the unit cell, lattice energy of VIC approximated to Madelung energy with point charges located at centers of defect and 0D species in vacuum. Energy required to transfer electrons between isolated defect and 0D species was derived from their charge and energy in VIC. Finite size effect was removed by extrapolation to infinite expanding substrate. Our results of the presented examples were in accord with reported theoretical and experimental results. The method evaded from consideration of material's dielectric constants, and had no need of improvement of DFT program. We expected its wide use to calculations of ion doping and adsorption on material's surface.

**Acknowledgments**

The work was supported by the NSFC Nos. 11274179 and 21203099, by National 973 projects of China with 2012CB921900, by Research Program for Advanced and Applied Technology of Tianjin (13JCYBJC36800). We appreciate the supports from the National Super-Computing Center at Tianjin and Guangzhou.

# References

[1]. I. Dabo, B. Kozinsky, N. E. Singh-Miller, and N. Marzari. Phys. Rev. B 77, 115139 (2008).

[2]. S. E. Taylor and F. Bruneval, Phys. Rev. B 84, 075155 (2011).

[3]. S. B. Zhang and A. Zunger, Phys. Rev. Lett. 77, 119 (1996).

[4]. H.-P. Komsa and A. Pasquarello, Phys. Rev. Lett. 110, 095505 (2013).

[5]. H.-P. Komsa, N. Berseneva, A. V. Krasheninnikov, and R. M. Nieminen, Phys. Rev. X 4, 031044 (2014).

[6]. D. Wang, D. Han, X. B. Li, S. Y. Xie, N. K. Chen, W. Q. Tian, D. West, H. B. Sun, and S. B. Zhang, Phys. Rev. Lett. 114, 196801 (2015).

[7]. S. Kim, K. J. Chang and J-S. Park, Phys. Rev. B 90, 085435 (2014).

[8]. K. Krishnaswamy, C. E. Dreyer, A. Janotti, and C. G. Van de Walle, Phys. Rev. B 92, 085420 (2015)

[9]. J. P. Perdew, K. Burke, and M. Ernzerhof, Phys. Rev. Lett. 77, 3865 (1996).

[10]. G. Kresse and D. Joubert, Phys. Rev. B 59, 1758 (1999).

[11]. W. Tang, E. Sanville, and G. Henkelman, J. Phys.: Condens. Matter 21, 084204 (2009).

[12]. http://www.jp-minerals.org/vesta/en/

[13]. C. G. Van de Walle and J. Neugebauer, J. Appl. Phys. 95, 3851 (2004).

[14]. J. Repp, G. Meyer, S. Paavilainen, F. E. Olsson, and M. Persson, Phys. Rev. Lett. 95, 225503 (2005).